\documentclass[twocolumn,secnumarabic,amssymb, amsmath, aps, prx]{revtex4-2}

\usepackage{graphicx}
\usepackage{natbib}
\usepackage[colorlinks,citecolor=blue,urlcolor=blue,hypertexnames=true,bookmarks=true]{hyperref}

\begin{document}
\onecolumngrid
{
\it
\hspace{3.75cm} Accepted in Soft Matter {\copyright} 2021 Royal Society of Chemistry
}
\vspace{-5pt}
\twocolumngrid
\title{Guided run-and-tumble active particles: wall accumulation and preferential deposition}%
\thanks{The work was performed while the author was at the Central University of
	Punjab. A version of the program code for run-and-tumble active particle
	simulations, in octave, is made public at: \href{https://gitlab.com/iamchamkor/bio3d}{ https://gitlab.com/iamchamkor/bio3d}.}
\author{Chamkor Singh}
\thanks{Email: iamchamkor@gmail.com}
\affiliation{Department of Physics, Central University of Punjab, Bathinda 151401, India\\}
\affiliation{Department of Chemical and Biological Engineering, Northwestern University, Evanston, IL 60208, USA}
\date{September 10, 2021}%
\begin{abstract}
Bacterial biofilms cost an enormous amount of resources in the health, medical, and industrial sectors. To understand early biofilm formation, beginning from planktonic states of active bacterial suspensions (such as {\it Escherichia coli}) to microcolonization, it is vital to study the mechanics of accumulation near surfaces and subsequent deposition. In this study, analytical expressions for the mean orientation, density and angular distributions, and deposition rates in such bacterial suspensions are derived, with and without the effects of external guiding or taxis fields. Simulations of confined active particles, using the run-and-tumble statistics from well-established three-dimensional tracking experiments and a preferential sticking probability model for deposition, closely verify the derived mean orientation, density profiles, angular distributions, and deposition rates. It is found that the size distribution of deposited microcolonies remains unaffected when guiding fields are applied, however, the pair correlation function of deposited structures relatively spreads out. The factor behind the changes in the accumulation patterns, and the changes in the architecture of deposited biomass, turns out to be an asymmetrical rotational drift caused by the guiding fields, and is an important physical mechanism behind the organization in confined active particle suspensions.       
\end{abstract}
\maketitle
%
%
%
%
%
\section{Introduction}
Multicellular microbial colonization or biofilm formation on surfaces affects numerous biological processes, medical technologies, supplies of drinking water~\cite{prest2016biological}, contamination of food surfaces~\cite{burnett2000attachment}, and can be a primary cause of certain diseases. The beginning of a cell's adhesion to a surface is marked by its transition from a fluid suspended (planktonic) state to a reversibly or irreversibly attached (sessile or deposited) state. Before reaching a surface, the transport of planktonic cells is influenced by factors like cell's active motility, advection by the surrounding fluid, its translational and rotational diffusion caused by surrounding fluid molecules, translational and rotational drifts due to different taxis mechanisms~\cite{lovely1975statistical,codutti2019chemotaxis, rossy2019cellular, guasto2012fluid}, characterstics of motion trajectories~\cite{ariel2015swarming, berg2004coli}, and so on. Above physical mechanisms play a role to bring the cells near surfaces where initial adhesion takes place, which over time converts to firm attachment via complex physio-chemical processes, bacterial surface developments, and competition between different forces e.g. Van der Waals, steric, and Coulomb~\cite{van1988bacterial,berne2018bacterial,palmer2007bacterial}. Eventually the aggregating cells form surface microcolonies, which mature and transform to biofilm~\cite{kimkes2020bacteria, sharma2014attachment,petrova2012sticky} exhibiting spatial and temporal patterns~\cite{thomen2020spatiotemporal}. The correspondence, collective effects and cooperation between cells are also complex aspects of early biofilm formation~\cite{lee2020social, paula2020dynamics, haagensen2015development}, in addition to competition for nutrients~\cite{xavier2009social}. Spatio-temporal patterns emerge in planktonic active suspensions, for instance, due to alignment~\cite{vicsek1995novel}, due to directional anisotropy of particles (i.e. a moving active particle is likely to encounter more neighbors in front than in back or on other sides)~\cite{pohl2016chemotaxis}, and even under isotropic repulsive interactions~\cite{fily2012athermal}. Physical processes associated with bacterial lifestyle are also altered by confined flows~\cite{conrad2018confined,hernandez2005transport,wioland2013confinement} and the nature of near-surface trajectories~\cite{lauga2006swimming,molaei2014failed}. Under certain conditions arising due to wall confinements, the cells may also exhibit accumulation near surfaces ~\cite{elgeti2013wall,sartori2018wall,li2011accumulation,li2009accumulation}. 
\begin{figure}
	\centering
	\includegraphics[width=1.0\linewidth]{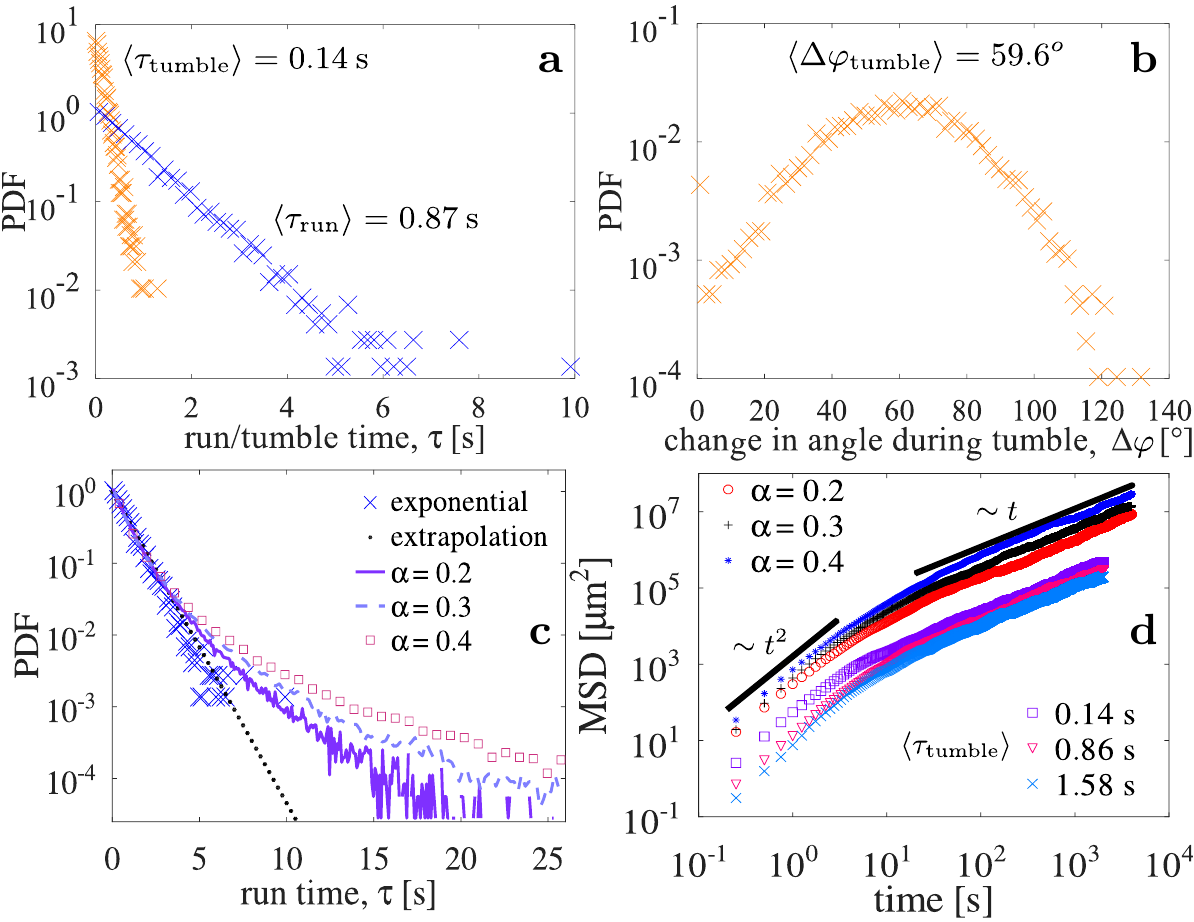}
	\caption{Probability distribution functions of (a) run and tumble durations, and of (b) tumble angles from simulations replicating the tracking experiments of~\citet{berg1972tracking}. {(c) Experiments by~\citet{korobkova2004molecular} and ~\citet{figueroa20203d} however indicate variability in running strategies in certain mutants resulting in heavy-tailed run-time distribution. Such run-time behavior is replicated by generalized Pareto distribution with a varying shape parameter $\alpha$ in the present simulations. (d) Mean square displacement (MSD), averaged over 100 realizations of an active particle executing run-and-tumble motion, both for exponentially distributed (with varying $\langle\tau_\mathrm{tumble}\rangle$) and heavy-tailed (with varying $\alpha$) run times. Although the ballistic to diffusive cross over time and the magnitude of MSD is slightly altered, the dynamics remains diffusive at long time scales for both types of run-time distributions.}}
	\label{fig_inputs}
\end{figure}
It is expected that near-wall accumulation of bacterial cells can alter early biofilm formation. In addition, the planktonic cells have shown preference to attach to an already existing colony, rather than attaching to a bare surface~\cite{grinberg2019bacterial} -- the process termed as preferential deposition. Keeping these two important and subtle aspects in mind, a theoretical and numerical framework is developed in this study to predict (i) accumulation profiles of planktonic cells near a surface, and (ii) architecture of microcolonies resulting from preferential deposition of the cells, with and without external guiding mechanisms. In addition to thermal fluctuations, individual {\it E. coli} trajectories consist of "run" (forward motility) and "tumble" (directional change) events, which are further associated with counterclockwise (run) and clockwise (tumble) motion of the bio-motors rotating the helical flagella of the bacterium~\cite{korobkova2004molecular,figueroa20203d}. For simplistic description, the run events can be described by following an exponential run-time distribution~\cite{berg1972tracking}, however over time, more intricacies and variabilities associated with the running process are being revealed~\cite{korobkova2006hidden,wang2017non,korobkova2004molecular,figueroa20203d,emonet2008relationship,waite2018behavioral,tu2005white}. \hspace{-5pt}The numerical strategy in this study consists of run-and-tumble dynamics utilizing the statistics from three-dimensional tracking experiments of~\citet{berg1972tracking}, as well as the variability in running behavior from more advanced experiments e.g. by~\citet{korobkova2004molecular} and ~\citet{figueroa20203d}.
In addition, the simulations use Rodrigues' relation for reorientational dynamics, a twitching interaction force to model excluded volume effects in the planktonic as well as the deposited state, and a rotational drift if an external guiding field is applied. 
A kinetic integral is developed to predict deposition rates. Under guiding fields, Fokker-Planck descriptions are used to predict the orientational and configurational distributions of particles in periodic and wall-bounded systems. The kinetic integral is then updated to include the effect of the external guiding field.   
\section{Computations}
Simulations involve three-dimensional motion of $N$ identical active particles, where each particle is either in a state of motion with a constant speed $v_o$ (run), or in a state of changing its direction of motion (tumble). A bacterium executes either a run or a tumble at a given time, and cannot be in both states simultaneously. In addition, it undergoes translational as well as rotational diffusion. The direction of motion changes, in a time step $\Delta t$, by an angle $\Delta\varphi$, which includes contributions of tumble events and rotational diffusion
\begin{equation}
\Delta\varphi = \sigma (1-s)\Delta\varphi_\mathrm{tumble} 
+
\sigma[
\sqrt{{2k_\mathrm{B}T}/{\gamma_\mathrm{r}}} \:\:{\boldmath \xi_\mathrm{r}}
] \Delta t.
\label{eq_delphi}
\end{equation}
If a bacterium is in a running state ($s=1$), its orientation changes only due to rotational diffusion, and in a tumbling state ($s=0$), it rotates with an additional angle $\Delta\varphi_\mathrm{tumble}$, at the end of the tumble duration. The change $\Delta\varphi_\mathrm{tumble}$ is chosen from a distribution, similar to one reported in three dimensional tracking experiments of~\citet{berg1972tracking} or in experiments by~\citet{figueroa20203d}. If a bacterium is already deposited on a surface ($\sigma=0$), there is no rotation and $\Delta\varphi=0$. In Eq.~\ref{eq_delphi}, $T$ is temperature of surrounding liquid, $k_\mathrm{B}$ is Boltzmann constant, $\gamma_\mathrm{r}$ is rotational friction coefficient, and $\xi_\mathrm{r}$ is white Gaussian noise with the properties $\langle\xi_\mathrm{r} (t) \rangle=0$ and $\langle\xi_\mathrm{r} (t) \xi_\mathrm{r}(t')\rangle = \delta(t-t')$, where $\delta$ is the Dirac delta. Numerically, $\xi_\mathrm{r}=w/\sqrt{\Delta t}$ where $w$ are machine generated discrete random numbers with zero mean and unit standard deviation. Knowing $\Delta\varphi$, the direction of motion, or the unit orientation vector $\mathbf{e}$ attached to the bacterium is rotated according to Rodrigues' relation       
\begin{equation}
\mathbf{e}^*= \sigma  [\mathbf{e} \cos \Delta\varphi + (\mathbf{n}\times\mathbf{e}) \sin \Delta\varphi],
\end{equation}
where $\mathbf{n}$ is a unit axis of rotation chosen randomly but with the constraint  $\mathbf{n}\cdot\mathbf{e}=0$, and ``$^*$" implies updated value. If a guiding torque $\mathbf{T}$ is also acting, then the direction of motion is changed by an additional angle $\Delta \varphi_2=[\mathbf{T}\cdot\mathbf{w}/\gamma_\mathrm{r}]\Delta t$ (details in section \ref{sec_accumulation}) about the axis of rotation $\mathbf{w}=\mathbf{T}/|\mathbf{T}|$. If $\mathbf{T\neq0}$ then bacterium reorients to $\mathbf{e}'= \sigma  [\mathbf{e}^* \cos \Delta\varphi_2 + (\mathbf{w}\times\mathbf{e}^*) \sin \Delta\varphi_2]$, and if $\mathbf{T=0}$ then $\mathbf{e}'= \mathbf{e}^*$~\footnote{For more details and help, refer to a version of the program code for run-and-tumble active particle simulations, and description therein, public at: \href{https://gitlab.com/iamchamkor/bio3d}{https://gitlab.com/iamchamkor/bio3d}.}. 
After reorientation of the bacterium, it is translated according to
\begin{equation}
\mathbf{r}' = \mathbf{r} +  \left[ \sigma\sqrt{{2k_\mathrm{B}T}/{\gamma_\mathrm{t}}} \: \boldsymbol{\xi}_\mathrm{t} + \sigma s v_o \mathbf{e}'  + \mathbf{F}_\mathrm{twitch}/ \gamma_\mathrm{t}  \right] \Delta t.
\end{equation}
Again if the bacterium is already deposited ($\sigma=0$), there is no translation except twitching. In tumble state $s=0$, the bacterium has only translational diffusion, or a twitching step ($\mathbf{F}_\mathrm{twitch} \Delta t/ \gamma_\mathrm{t}$) if it overlaps with other bacteria. The twitching force $\mathbf{F}_\mathrm{twitch}$ takes care of the excluded volume and acts only between overlapping bacteria. It also effectively models the twitching motility of bacteria during the deposited state. It is applied such that if two bacteria overlap by an amount $\delta$, they are moved apart by exactly the same distance $\delta$ in a time step $\Delta t$. The following relation executes this
\begin{equation}
\mathbf{F}_\mathrm{twitch} = \frac{\gamma_\mathrm{t}\delta}{2\Delta t}\:\mathbf{u},  
\label{eq_twitching}	
\end{equation}
for $\delta>0$, and $\mathbf{F}_\mathrm{twitch} =0$ if $\delta<=0$. Here $\mathbf{u}$ is unit normal pointing between centers of overlapping bacteria, and chosen such that $\mathbf{F}_\mathrm{twitch}$ is repulsive. The components of noise vector $\boldsymbol{\xi}_\mathrm{t}$ have same properties as $\boldmath{\xi}_\mathrm{r}$, and the two are uncorrelated. For a given bacterium, the duration for a run ($\tau_\mathrm{run}$) or a tumble ($\tau_\mathrm{tumble}$) are chosen from experimental distributions~\cite{berg1972tracking,figueroa20203d,berg2004coli}. Realizations and motion statistics of the numerical model are shown in Fig.~\ref{fig_inputs}. 
\begin{figure}[b!]
	\centering
	\includegraphics[width=0.99\linewidth]{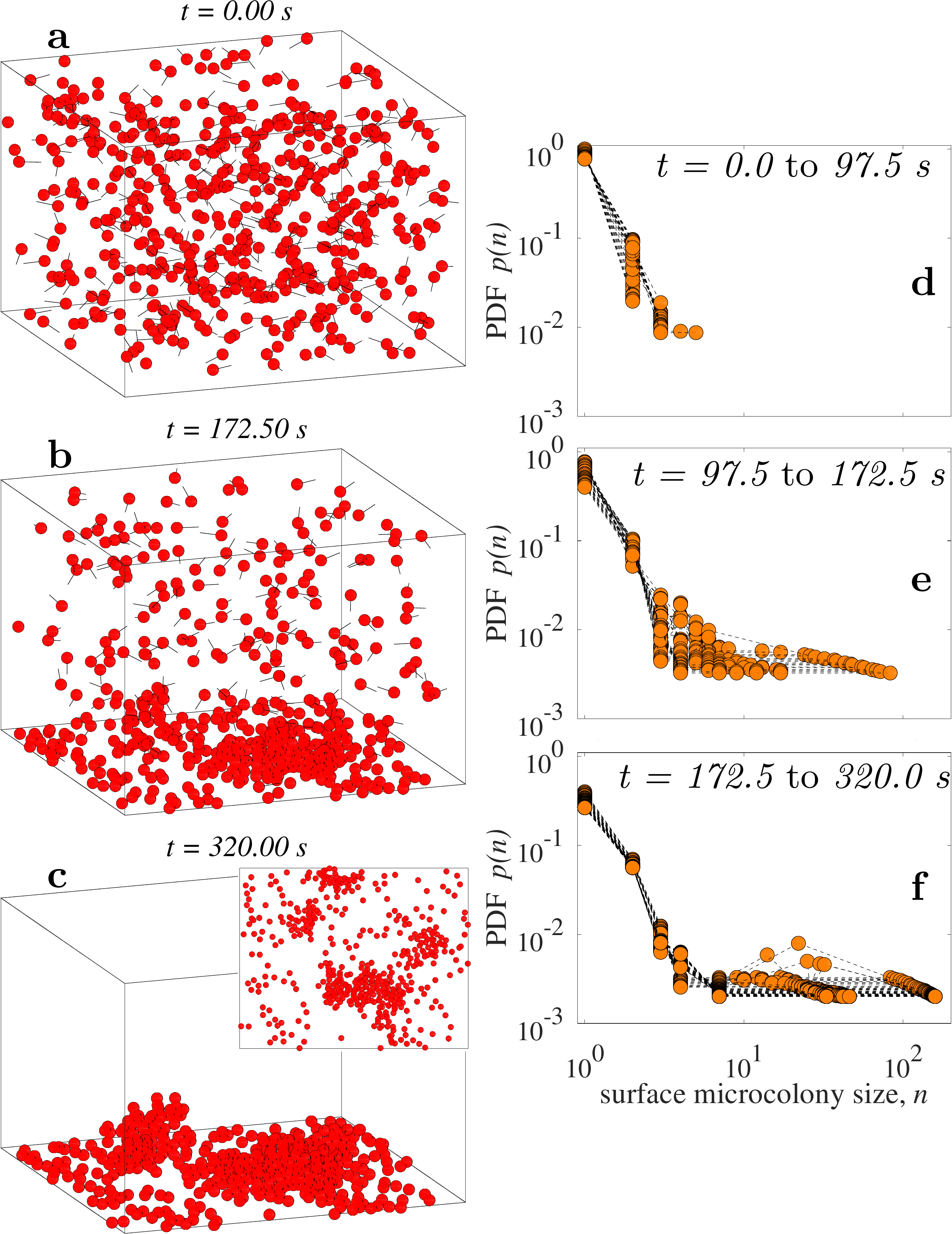}
	\caption{(a-c) Snapshots from a typical simulation of run-and-tumble active suspension, depositing on the bottom surface $[z=-L_z/2]$, in the absence of any guiding torque. The line segments in the suspended state indicate particle orientations. (c, inset) Top view of the deposited particles. (d-f) Development of the probability distribution of sizes of the microcolonies due to preferential deposition, with time. While the small colonies ($n=1$ or $2$) are more likely, colonies with sizes greater than $n\approx5$ are rare but are almost equally probable.}
	\label{fig_scatter}
\end{figure}
When a bacterium reaches the bottom surface ($z=-L_z/2$), it may attach to the surface with a preferential probability. Here preferential means that the attachment probability increases with the increasing size of a colony with which the bacterium comes into contact and seeks attachment~\cite{grinberg2019bacterial}. In the present study, the preferential deposition of bacteria onto microcolonies is modeled using a sticking probability (or probability of attachment) which depends on the size $n$ of colonies with which bacteria come in contact, and is given by  
\begin{equation}
P = \frac{ P_\mathrm{s} P_\mathrm{c} }{P_\mathrm{s} +(P_\mathrm{c}-P_\mathrm{s})\exp (-\lambda n) }
.
\label{eq_p}
\end{equation}
For a single bacterium which comes in contact with the surface, $n=0$ and $P=P_\mathrm{s}$. As $n$ increases, $P$ approaches $P_\mathrm{c}$ logistically. The model have three parameters $P_\mathrm{s},\:P_\mathrm{c}$, and $\lambda$. It is assumed that a bacterium which comes in contact with the surface has a rare chance of attachment ($P_\mathrm{s}\sim 0.1\%$). If it hits a colony with size $n$, the chance of attachment is given by Eq.~\ref{eq_p}. The highest possible probability in simulations is limited to $P_\mathrm{c}\rightarrow 40\%$ as colony size $n\rightarrow \infty$. Other fixed parameters in the present study are the domain size $L_x=L_y=L_z=30\:\mu\mathrm{m}$, $\Delta t=0.025$ s, volume fraction $\sim 1\%-5\%$, cell diameter $1\:\mu\mathrm{m}$, translational and rotational diffusion coefficients $k_\mathrm{B}T/\gamma_\mathrm{t}=k_\mathrm{B}T/\gamma_\mathrm{r}=0.05$, particle numbers $500-5000$, and swimming speed $v_o=10\:\mu\mathrm{m}\:\mathrm{s}^{-1}$. 
Mean tumble time $\langle \tau_\mathrm{tumble}\rangle=0.14$ s, mean angular change during tumble events $\langle\Delta \varphi_\mathrm{tumble}\rangle\approx60^o$, and $\langle\Delta \varphi_\mathrm{tumble}^2\rangle^{1/2}\approx20^o$ are adopted from experiments of~\citet{berg1972tracking} [Fig.~\ref{fig_inputs} (b)]. 
{ The boundary conditions in the $x$ and $y$ directions are periodic, while the particles are reflected back, or deposited, on the walls in $z$ direction.} 

{ For the run events, two type of strategies are considered. In the first case, exponentially distributed run times with mean run time $\langle \tau_\mathrm{run}\rangle=0.86$ s are chosen~\cite{berg1972tracking} [Fig.~\ref{fig_inputs} (a)]. More recent and advanced experiments, e.g. by~\citet{korobkova2004molecular} or~\citet{figueroa20203d}, however report that the running strategies in certain mutant bacterial populations exhibit behavioral variability and heavy-tailed run-time distributions. Therefore a second running strategy mimicking such effects is also considered in the present simulations. In this case, the run times are chosen from a generalized heavy-tailed Pareto distribution with a scale parameter equal to $\langle\tau_\mathrm{run}\rangle$ and a varying shape parameter $\alpha$ [Fig.~\ref{fig_inputs} (c)]. For $\alpha=0$, exponential distribution is retrieved. 
	The choice of heavy-tailed run times cause a fundamental change in the nature of individual particle trajectories. Although the dynamics remains diffusive at long time scales for both types of run-time distributions, the ballistic to diffusive cross over time and the magnitude of mean square displacement (MSD) is slightly altered. This is depicted in Fig.~\ref{fig_inputs} (d). The MSD is averaged over 100 realizations of an active particle executing run-and-tumble motion, both for exponentially distributed (with varying $\langle\tau_\mathrm{tumble}\rangle$) and heavy-tailed (with varying $\alpha$) run times. 
	A typical simulation of particles undergoing a transition from planktonic to preferentially deposited states is shown in Fig.~\ref{fig_scatter} (a-c). The microcolonies so formed can have a certain height [Fig.~\ref{fig_scatter} (c)], and size distribution $p(n)$ [Fig.~\ref{fig_scatter} (d-f)], where $n$ is the microcolony size i.e. the number of bacteria in a given microcolony. The probability density $p(n)$ nearly follows a power law for small microcolonies $n\sim 5$, however, it is rather flat for larger size microcolonies.  

	A framework for the rate of deposition of unguided particles is described in the next section, which is further developed for the case of external guiding field application in section~\ref{sec_dep_guided_particles}, after considering insights from the Fokker-Planck descriptions of guided motion in section~\ref{sec_accumulation}. A detailed discussion on the overall results from the study follows thereafter in section~\ref{sec_discussion}.        
}

\section{Kinetics of deposition: unguided particles}
Consider the volumetric number density near the bottom surface $z=-L_z/2$ is $\rho(\mathbf{r}_\mathrm{b})$. Then the number of particles in a differential volume $d\mathbf{r}_\mathrm{b}$ placed at the bottom surface is
$
\rho(\mathbf{r}_\mathrm{b}) d\mathbf{r}_\mathrm{b}
\label{eq_rho}
$
. Not all of the particles in this volume are oriented, and moving, towards the bottom surface. If $f(\mathbf{e})$ is the orientational distribution function, then the probability that a particle with orientation $\mathbf{e}$ around $d\mathbf{e}$ is approaching the bottom surface, will be $
f(\mathbf{e}) d\mathbf{e}\:\Theta(-\mathbf{e}\cdot \mathbf{n})
\label{eq_e}
$. Here $\mathbf{n}$ is the unit outward normal at the bottom surface, and $\Theta$ is the Heaviside function. Also, for a small time period $d t$, only the particles which are within the differential volume $
d\mathbf{r}_\mathrm{b} = |v_o\:\mathbf{e}\cdot\mathbf{n}\:ds| dt  
\label{eq_dr}
$ are available to reach the bottom surface. Here $ds$ is the differential surface area at the bottom surface, and thus $|v_o\:\mathbf{e}\cdot\mathbf{n}| dt$ is the height of the differential volume $d\mathbf{r}_\mathrm{b}$. If a particle touches the surface, it is deposited to the surface with a sticking probability $P$ [Eq.~\ref{eq_p}]. Combining 
the above, the rate of particles getting deposited on the bottom surface is
\begin{equation}
\frac{dN_\mathrm{dep}}{dt}=\int_s ds \int_{\mathbf{e}} d\mathbf{e}\:\:
P
\rho(\mathbf{r}_\mathrm{b}) |v_o\:\mathbf{e}\cdot\mathbf{n}| 
f(\mathbf{e})\:\Theta(-\mathbf{e}\cdot \mathbf{n}).
\label{eq_dNdt_int}
\end{equation}
The integral can be simplified assuming a homogeneous number density $(N-N_\mathrm{dep})/(L_xL_yL_z)$ of particles available for deposition, the surface area $\int ds$ available for deposition at the bottom equals to $L_x L_y$, and assuming a uniform $f(\mathbf{e})$. Then the deposition rate is simplified to
\begin{equation}
\frac{dN_\mathrm{dep}}{dt}=
P\frac{ v_o (N-N_\mathrm{dep})}{L_z}
\frac{1}{4\pi} 
I_\mathbf{e},
\label{eq_dNdt}
\end{equation}
where $I_\mathbf{e}=
\int_{\mathbf{e}} 
d\mathbf{e}
|\mathbf{e}\cdot\mathbf{n}|
\Theta(-\mathbf{e}\cdot \mathbf{n})
=\pi$. The solution of this equation is straightforward and reads
\begin{equation}
N_\mathrm{dep}(t)=N-N\exp \left[
-\frac{Pv_o}{4 L_z}\: t
\right].
\label{eq_Ndep_Pconst}
\end{equation}
Notice that a deposition time scale $\tau_\mathrm{dep}=\frac{4L_z}{Pv_o}$ emerges from the above solution. $\tau_\mathrm{dep} \gg 1$ would imply that the particles take longer to get deposited. A smaller confinement height $L_z$, larger propulsion speed $v_o$, and/or larger sticking probability help accelerating deposition. Also from Eq.~\ref{eq_Ndep_Pconst}, the half-time of deposition is $\frac{ 4 L_z \ln(2)}{P v_o}$ which is $\approx 0.7\:\tau_\mathrm{dep}$, meaning that $70\%$ of the time is consumed in depositing first half of the particle population. The second half is deposited relatively fast. The solution of Eq.~\ref{eq_dNdt} with $P$ from Eq.~\ref{eq_p}, however, is more involved and in implicit form it reads

\begin{figure}[!b]
	\centering
	\includegraphics[width=0.7\linewidth]{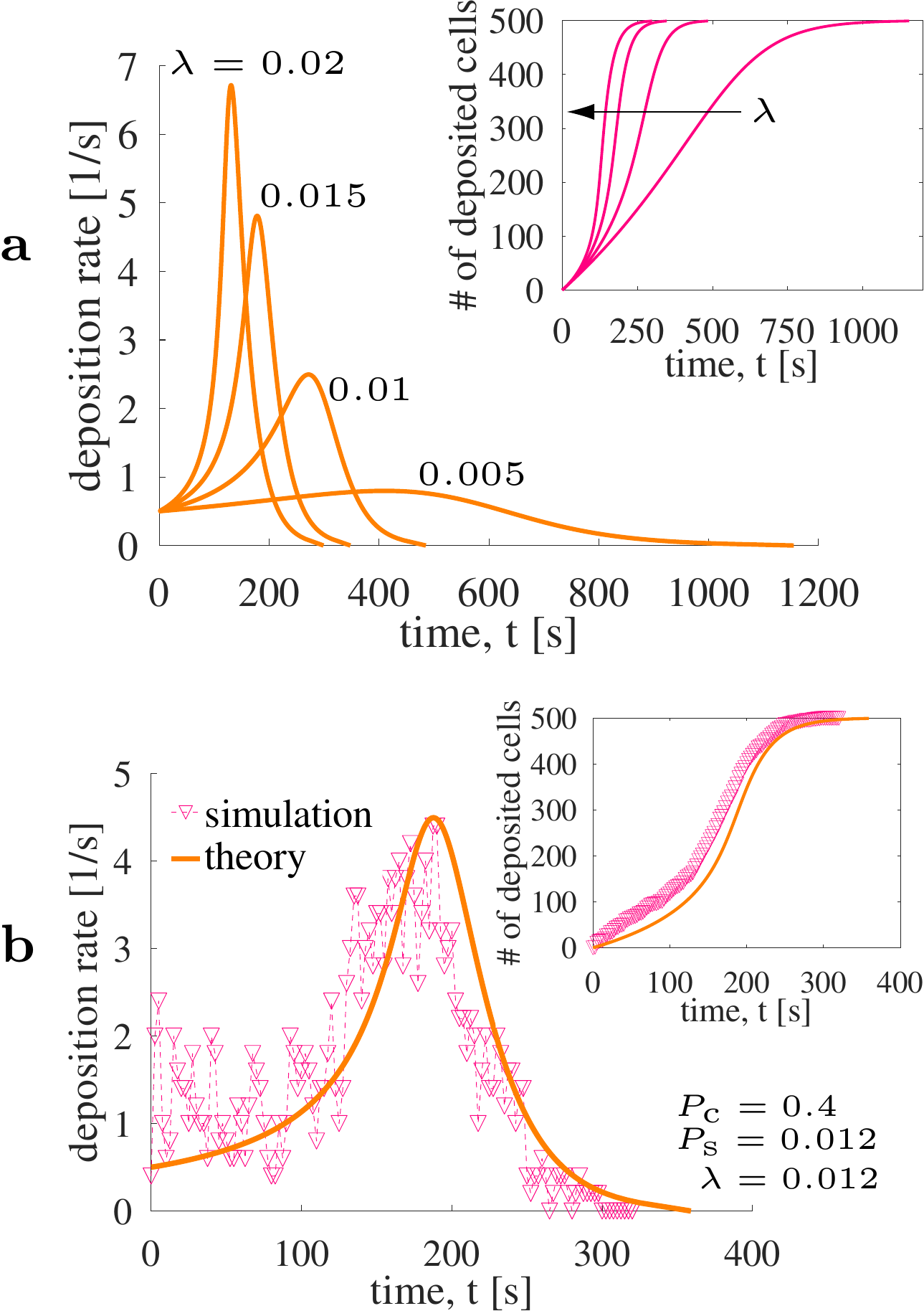}
	\caption{(a) Deposition rate $dN_\mathrm{dep}/dt$, and number of deposited particles $N_\mathrm{dep}$ with time for different $\lambda$, predicted by Eq.~\ref{eq_soln_Ei_1}. (b) Comparison with simulation.}
	\label{fig_solution}
\end{figure}
\begin{figure}[!t]
	\centering
	\includegraphics[width=1.0\linewidth]{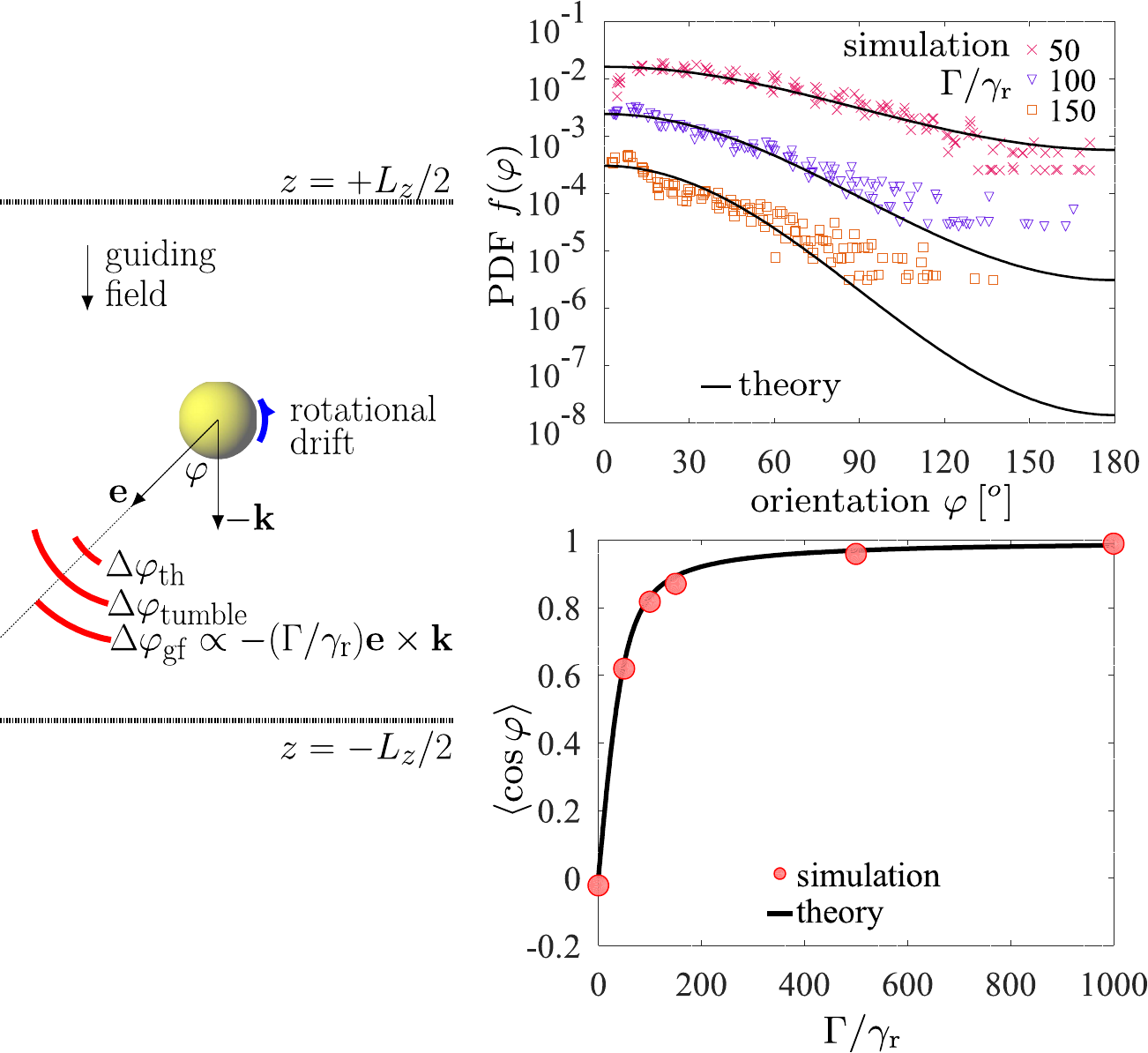}
	\caption{A guiding field or taxis mechanism, induces a rotational drift in particle suspension in addition to thermal noise and run-and-tumble events. Therefore it changes the angular distribution and inducing a mean flow. (Left) Schematic for change in orientation of an active particle due to thermal noise $[\Delta\varphi_\mathrm{th}=\Delta t \sqrt{2 k_\mathrm{B} T/\gamma_\mathrm{r}} \xi_\mathrm{r}]$, tumbling $[\Delta\varphi_\mathrm{tumble}]$, and rotational drift $[\Delta\varphi_\mathrm{gf}=-\Delta t \:\Gamma \sin \varphi/\gamma_\mathrm{r}]$. The thermal and tumble noises are symmetric about the direction of motion, while rotational drift is biased towards guiding field direction. (Top-right) Probability density of angle between direction of motion and -ve $z$ direction. The PDF peaks towards smaller angles as the guiding strength is increased. The data for $\Gamma/\gamma_\mathrm{r}=100,\:150$ is shifted vertically for clarity. (Bottom-right) Increase of guiding torque strength increases the order parameter for mean orientation. Symbols are for simulations and lines are stationary solutions of Fokker-Planck equation.}
	\label{fig_fokker_planck}
\end{figure}
\begin{widetext}
\begin{equation}
-P_\mathrm{s}\ln \frac{N-N_\mathrm{dep}}{N}
-(P_\mathrm{c}-P_\mathrm{s})e^{-\lambda N} 
[Ei(\lambda N-\lambda N_\mathrm{dep})-Ei(\lambda N)]
= \frac{v_oP_\mathrm{c}P_\mathrm{s}}{4 L_z} t,
\label{eq_soln_Ei_1}
\end{equation}
\end{widetext}
where $Ei(x)=-\int_{-x}^{\infty} e^{-b} b^{-1}db$ is the exponential integral. This solution is shown in Fig.~\ref{fig_solution} (a) for different values of $\lambda$ (which affects the change in the sticking probability $P$, from $P_\mathrm{s}$ to $P_\mathrm{c}$, as the size of microcolonies increase, Eq.~\ref{eq_p}). A typical case is compared with simulation in Fig.~\ref{fig_solution} (b). Assumptions of uniform spatial and orientational distributions in solution Eq.~\ref{eq_soln_Ei_1} work well in the absence of any guiding field or taxis mechanism, e.g. chemotaxis, galvanotaxis, or magnetotaxis. The presence of such guiding mechanisms adds to the physical complexity, and the distributions need to be tweaked. In the following, a Fokker-Planck description is developed for spatial and angular distributions to update the solution in Eq.~\ref{eq_soln_Ei_1}, accounting for the particle accumulation near the depositing surface and the subsequent effect on deposition rate. In the numerical model, it implies the direct application of non-zero $\mathbf{T}$, however in the analytical model, it needs systematic derivations of the spatial and angular distributions, addressed in the following.
\vspace{-7.9pt}  
\section{Accumulation of guided particles\label{sec_accumulation}}
From Eq.~\ref{eq_dNdt_int} it is learned that the deposition rate may be controlled by tweaking particle density $\rho(\mathbf{r})$ near the wall, orientation distribution $f(\mathbf{e})$, and/or the sticking probability $P$. To achieve this in practice, different taxis/guiding mechanisms, such as chemotaxis, galvanotaxis, or magnetotaxis etc. can be used. Let us consider that an applied taxis field in a direction $\mathbf{d}$ tends to align the bacterium along $\mathbf{d}$ and thus results in a torque $\mathbf{T}=\Gamma \mathbf{e}\times\textbf{d}$. Here $\Gamma$ is the strength. In the present case, $\mathbf{d}$ is considered along the -ve $z$ direction i.e. $-\mathbf{k}$. If the angle between particle orientation $\mathbf{e}$ and the direction $-\mathbf{k}$ is $\varphi$ [Fig.~\ref{fig_fokker_planck}], then $\mathbf{T}=\Gamma \sin\varphi \:\mathbf{w}$, with $\mathbf{w}=-\mathbf{e}\times\textbf{k}/|-\mathbf{e}\times\textbf{k}|$. Considering a symmetric situation about $-\mathbf{k}$, and introducing an effective temperature $T_\mathrm{eff}$ which lumps together the tumbling and rotational diffusion, Eq.~\ref{eq_delphi} is modified to
\begin{equation}
\Delta\varphi =
\sigma[
\sqrt{{2k_\mathrm{B}T_\mathrm{eff}}/{\gamma_\mathrm{r}}} \:\:{\boldmath \xi_\mathrm{r}}
-
{\Gamma} \sin \varphi/{\gamma_\mathrm{r}}
] 
\Delta t.
\label{eq_delphi_gamma}
\end{equation}
Again already deposited particles ($\sigma=0$) undergo no rotation. The guiding torque introduces an asymmetry in steps $\Delta\varphi$~[Fig.~\ref{fig_fokker_planck}] giving rise to an asymmetrical angular drift. The incorporation of guiding torque in the theory is described below, in angular as well as in configuration space.

\subsection{Fokker-Planck description: angular space}
The time evolution of $f(\varphi)$ is estimated assuming that the shifts in angle are Markovian, and satisfy the Chapman-Kolmogorov equation $
f(\varphi,t+\Delta t)=\int_{\Delta\varphi} f(\varphi-\Delta\varphi,t)q(\varphi-\Delta\varphi,\Delta\varphi)d(\Delta\varphi)
$, where $q(\varphi-\Delta\varphi,\Delta\varphi)$ is the probability density of a particle with orientation $\varphi-\Delta\varphi$ to have a jump $\Delta\varphi$. Using Taylor expansions, the equation can be transformed to the Fokker-Planck equation
\begin{equation}
\frac{\partial f}{\partial t} = 
- \frac{\partial}{\partial \varphi}
D_1(\varphi)
f
+ \frac{\partial}{\partial \varphi}
D_2(\varphi)	
\frac{\partial f}{\partial \varphi},
\label{eq_fpequ}
\end{equation}
where the drift coefficient $D_1$ and the diffusion coefficient $D_2$, using Eq.~\ref{eq_delphi_gamma}, turn out to be $
D_1 = \frac{1}{\Delta t}\int_{\Delta\varphi} \Delta\varphi\:q(\varphi,\Delta\varphi)\:d(\Delta\varphi)=-{\Gamma}\sin \varphi/\gamma_\mathrm{r}
$, and $
D_2 = \frac{1}{2\Delta t}\int_{\Delta\varphi} \Delta\varphi^2\:q(\varphi,\Delta\varphi)\:d(\Delta\varphi)={k_\mathrm{B}T_\mathrm{eff}}/{\gamma_\mathrm{r}}
$, respectively. Eq.~\ref{eq_fpequ} has stationary ($\partial_t f=0$) solution~\cite{schienbein1994random} 
\begin{equation}
f(\varphi) = f_o \exp\left[
\frac{\Gamma\cos\varphi}{k_\mathrm{B}T_\mathrm{eff}}
\right].
\label{eq_fphi}
\end{equation}
where the normalization factor is calculated as
\begin{equation}
f_o^{-1} =\int_{0}^{2\pi}  \exp\left[
\frac{\Gamma\cos\varphi}{k_\mathrm{B}T_\mathrm{eff}}
\right] d\varphi = 2\pi I_0(\Gamma/k_\mathrm{B}T_\mathrm{eff}).
\end{equation}
Here $I_0$ is the modified Bessel function of first kind. For $\Gamma=0$ (no guiding field), the angular distribution reduces to a uniform one $f=1/2\pi$. Knowing the stationary angular distribution, the mean orientation of particles is
\begin{equation}
\langle \cos\varphi \rangle = \int_{0}^{2\pi} \cos\varphi\: f(\varphi) d\varphi
= \frac{I_1(\Gamma/k_\mathrm{B}T_\mathrm{eff})}
{I_0(\Gamma/k_\mathrm{B}T_\mathrm{eff})}.
\label{eq_cosphi}
\end{equation}
The results for angular distribution and mean orientation are compared with simulations in Fig.~\ref{fig_fokker_planck}. In the following, the above solutions in Eq.~\ref{eq_fphi}-\ref{eq_cosphi}, originally proposed by~\citet{schienbein1994random}, are further augmented with descriptions of density non-uniformity, near-surface accumulation, and subsequent effects on deposition rate.   
\vspace{-10pt}  
\subsection{Fokker-Planck description: configuration space}
\begin{figure}[!t]
	\centering
	\includegraphics[width=0.98\linewidth]{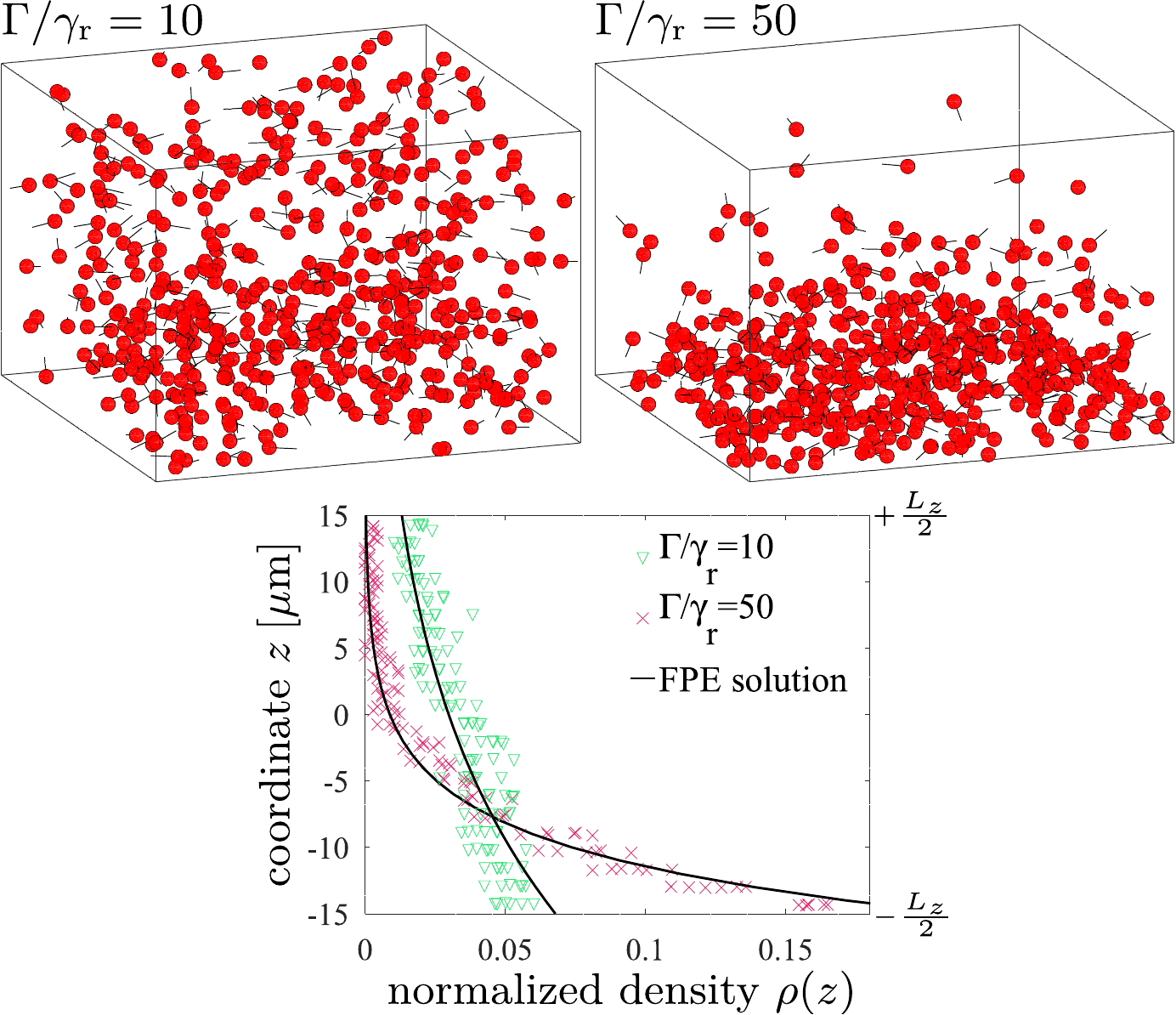}
	\caption{Accumulation of run-and-tumble active particle near a wall can be tweaked by changing the strength of guiding field or rotational drift. Normalized density variation along the vertical: (top) configurations from the simulations for $\Gamma/\gamma_\mathrm{r}=10$ and $50$ at $t=20\:\:\mathrm{s}$, when stationary profiles are already reached; (bottom) comparison of simulations with the prediction using Fokker-Planck equation [FPE, Eq.~\ref{eq_fpeq_rho},~\ref{eq_rho_solution}]. The accumulation of particles eventually leads to increased deposition rates and thus the deposition integral [Eq.~\ref{eq_dNdt_int}] needs to be updated incorporating this effect.}
	\label{fig_fokker_planck_rho}
\end{figure}

Let us consider the guiding field acting in $-\mathbf{k}$ direction leaves the density of particles in $x$ and $y$ directions homogeneous, and affects only $\rho(z)$. The particles experience translational drift in $-\mathbf{k}$ direction, in addition to translational thermal diffusion and tumble events. The resulting Fokker-Planck equation reads    
\begin{equation}
\frac{\partial \rho(z,t)}{\partial t} = 
- \frac{\partial}{\partial z}
(- v_o \langle \cos \varphi\rangle
)\rho
+ \frac{\partial}{\partial z}
\frac{k_\mathrm{B} T_\mathrm{eff} }{\gamma_\mathrm{t}}	
\frac{\partial \rho}{\partial z},
\label{eq_fpeq_rho}
\end{equation}
where mean orientation $\langle \cos \varphi\rangle$ is used in the translational drift coefficient to decouple the $f(\varphi)$ and $\rho(z)$ equations. The stationary solution to this equation reads
\begin{equation}
\rho(z) = \rho_o \exp\left[
-\frac{ v_o \gamma_\mathrm{t} \langle \cos \varphi\rangle  }{k_\mathrm{B} T_\mathrm{eff}}	z
\right],
\label{eq_rho_solution}
\end{equation}
with the normalization factor
\begin{equation}
\rho_o^{-1} = \int_{-Lz/2}^{Lz/2} \exp\left[
-\frac{ v_o \gamma_\mathrm{t} \langle \cos \varphi\rangle  }{k_\mathrm{B} T_\mathrm{eff}}	z
\right] dz.
\end{equation}
Particle configurations and the normalized density profiles for two different $\Gamma$ are shown in Fig.~\ref{fig_fokker_planck_rho}. The continuous solution Eq.~\ref{eq_rho_solution} closely agrees with the simulations, for a suspension even as small as $N=500$. The integration limits in the normalization factor $\rho_o^{-1}$ accounts for the bounding surfaces in $z$ direction. It was incredible to note that using these limits of $z$ was sufficient enough to replicate the simulations, even though the derivation of mean orientation and angular distribution did not include the information about the boundaries. Once density non-uniformity in $z$ direction is calculated, the deposition integral and its solution are updated in the following.
\vspace{-3.5pt}  
\section{Kinetics of deposition: guided particles~\label{sec_dep_guided_particles}}
Knowing the changes in angular and density distributions $f(\varphi)$ and $\rho(z)$ respectively under guiding field, the integral in Eq.~\ref{eq_dNdt_int} is updated to
\begin{widetext}
\begin{align}\nonumber
\frac{dN_\mathrm{dep}}{dt}
=
P
v_o
{\rho(-L_z/2)}
(N-N_\mathrm{dep})
\int_0^{2\pi}
d\theta
\int_0^{\pi}
d\varphi 
\times		
\sin\varphi
\:|\cos\varphi| 
\:f(\varphi) 
\:\Theta(-\cos\varphi)
\\\nonumber
=P
v_o
(N-N_\mathrm{dep})
\:2\pi
f_o
\rho_o
\exp\left[ \frac{v_o \gamma_\mathrm{t} L_z \langle\cos \varphi\rangle}{2k_\mathrm{B}T_\mathrm{eff}} \right]\times
\frac{1-\exp a + a \exp a}{a^2},
\end{align}
implicit solution of which reads
\begin{align}\nonumber
-P_\mathrm{s}\ln \frac{N-N_\mathrm{dep}}{N}
-(P_\mathrm{c}-P_\mathrm{s})e^{-\lambda N}
[Ei(\lambda N-\lambda N_\mathrm{dep})-Ei(\lambda N)]\\
\!=\! 
2\pi
v_o
f_o
\rho_o
\exp\!\left[ \frac{v_o \gamma_\mathrm{t} L_z \langle\cos \varphi\rangle}{2k_\mathrm{B}T_\mathrm{eff}} \right]\!\!\times
\frac{1\!-\!\exp a \!+\! a \exp a}{a^2} P_\mathrm{c}P_\mathrm{s} t,
\label{eq_soln_Ei}
\end{align}
\end{widetext}
where $a=\Gamma/k_\mathrm{B}T_\mathrm{eff}$, and $Ei(x)=-\int_{-x}^{\infty} e^{-b} b^{-1}db$ is the exponential integral. 
The solution is same to Eq.~\ref{eq_soln_Ei_1} except the fact that $v_o/4L_z$ on the right hand side of Eq.~\ref{eq_soln_Ei_1} is now replaced. This effect of guiding field on the deposition rate, together with simulations is plotted in Fig~\ref{fig_fokker_planck_dNdt}. Notice that the left hand side of Eq.~\ref{eq_soln_Ei} is same as Eq.~\ref{eq_soln_Ei_1}, however, in contrast to Eq.~\ref{eq_soln_Ei_1}, the deposition time scale emerging from the right hand side of Eq.~\ref{eq_soln_Ei} is now more involved, and includes guiding torque strength $\Gamma$, effective translational diffusion $k_\mathrm{B}T_\mathrm{eff}/\gamma_\mathrm{t}$, and mean orientation of particles $\langle\cos\varphi\rangle$ from Eq.~\ref{eq_cosphi}.         

\vspace{-7pt}  
\section{Discussion~\label{sec_discussion}}
Present theory based on the Fokker-Planck equations, and the simulations, depict that a taxis mechanism or guiding field induces an asymmetric rotational drift which eventually determines how the active particles are distributed in angular and configuration spaces [Eq.~\ref{eq_fphi},~\ref{eq_rho_solution} and Fig.~\ref{fig_fokker_planck},~\ref{fig_fokker_planck_rho}]. Understanding this physical mechanism is essential to the study of the accumulation of active suspensions near surfaces, and subsequently their deposition. Conversely, it can also be deduced that the application of reversed guiding fields can help reduce biomass deposition. Also, the tweaking of configurational and angular distributions in a controlled fashion can be an important parameter while designing different taxis devices, either to reduce or to enhance biomass buildup during early biofilm formation. The kinetic integral developed in this study [Eq.~\ref{eq_dNdt_int}], and the Fokker-Planck descriptions of configurational and angular distributions, closely predict the density variations due to wall accumulation as well as biomass deposition rates. The angular and translational motions can be considered independent of each other in the Fokker-Planck descriptions. In fact when $\langle \cos\varphi\rangle$ calculated in Eq.~\ref{eq_cosphi} for an extended system, or for periodic boundary conditions, is used to approximate the translational drift in the $\rho(z)$ equation~\ref{eq_fpeq_rho}, provides accurate results for $\rho$ profiles [Fig.~\ref{fig_fokker_planck_rho}]. It is however experienced that increasing $\Gamma/\gamma_\mathrm{r}$ leads to more dominant boundary effects, and increased deviation of theoretical $f(\varphi)$ and $\rho(z)$ predictions from the numerical predictions. 
\begin{figure}[b!]
	\centering
	\includegraphics[width=0.98\linewidth]{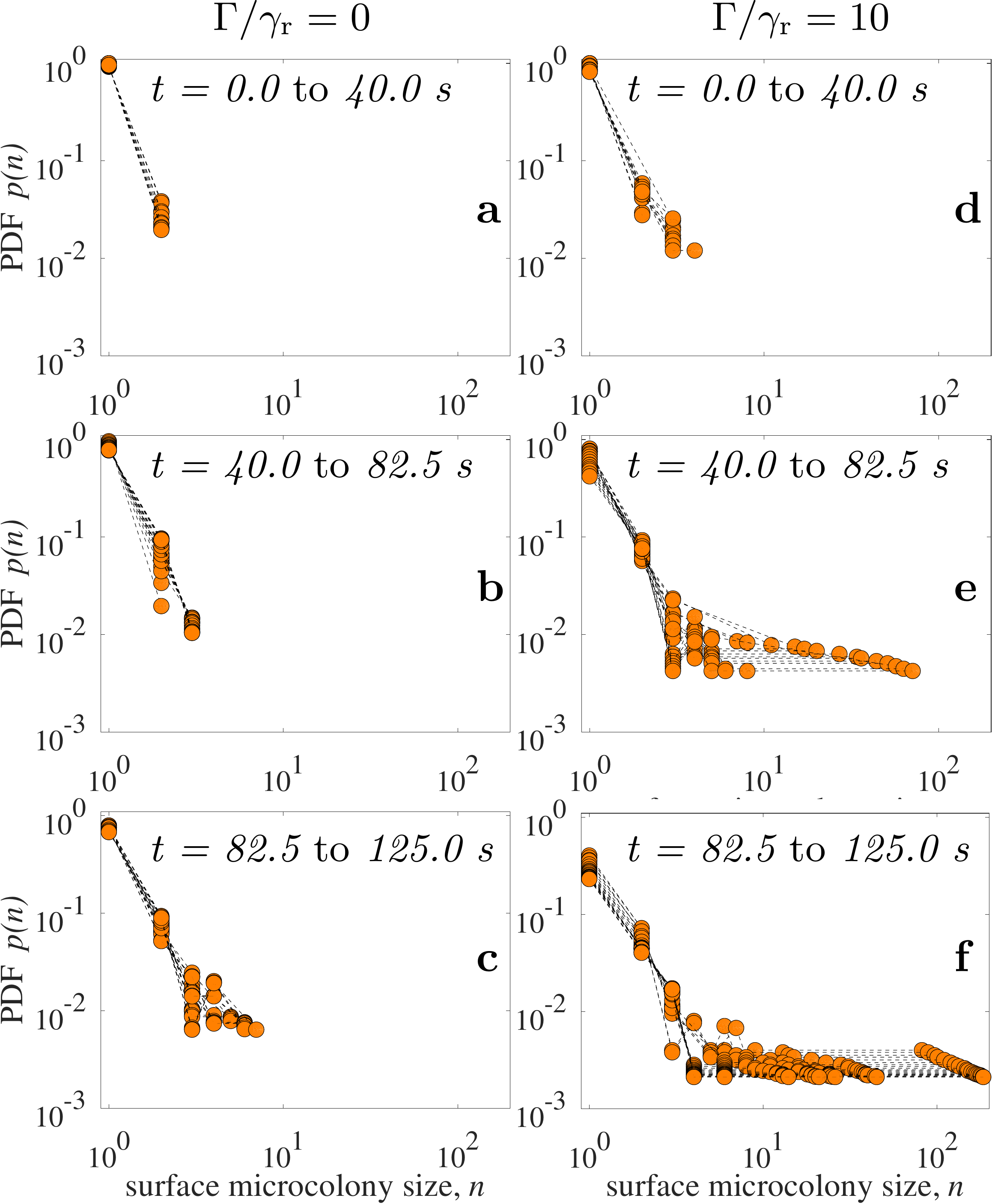}
	\caption{The size distribution of microcolonies matures relatively faster on increasing the guiding torque strength $\Gamma$ [panel (a-c) vs. panel (d-f)]. However, the effect on the eventual distribution -- obtained after deposition of all the particles -- is negligible (by comparing panel (f) for $\Gamma/\gamma_\mathrm{r}=10$ with Fig.~\ref{fig_scatter} (f) for $\Gamma/\gamma_\mathrm{r}=0$). This observation, combined with the result for pair correlation function in Fig.~\ref{fig_fokker_planck_dNdt}, signifies that the rotational drift arising due to the application of guiding field tends to alter the architecture of the deposited microcolonies and changes the deposition rate, without significantly influencing the eventual size distribution.}
	\label{fig_sizedist-combined-r00-vs-r10}
\end{figure}
\begin{figure}[t!]
	\centering
	\includegraphics[width=1.0\linewidth]{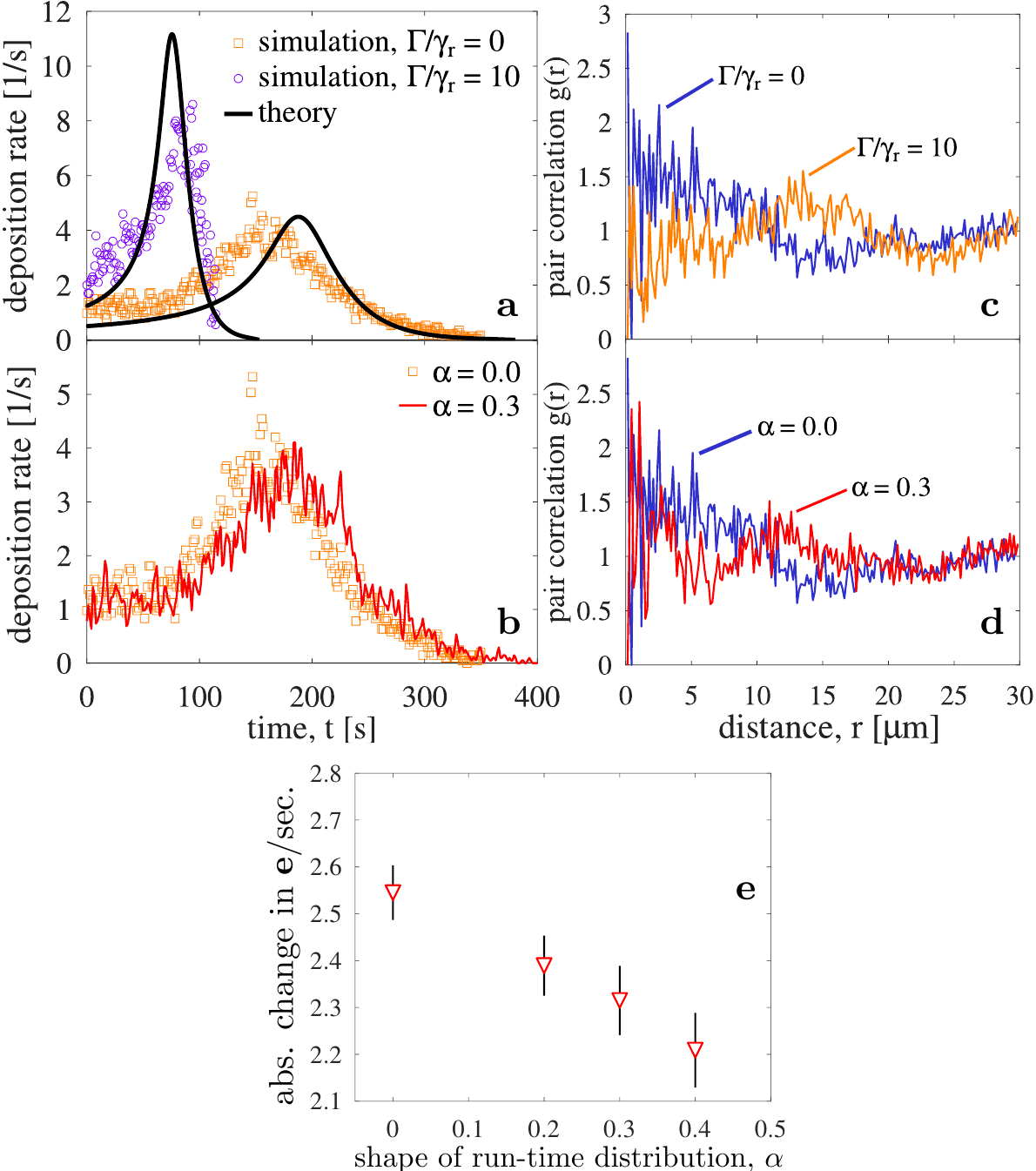}
	\caption{(a) Tweaking the guiding torque strength $\Gamma$ changes the dynamics of particle deposition rate, and is predicted closely by Eq.~\ref{eq_soln_Ei}. {(b) Comparison of deposition rate when the run-time distribution is changed from an exponential ($\alpha=0$) to a heavy-tailed one ($\alpha=0.3$). The pair correlation function (computed by first projecting the particles on the depositing plane and then computing the 2D pair-correlation) is affected upon increasing $\Gamma$ (c), or upon increasing $\alpha$ (d). This indicates that although the deposition rate is altered significantly only due to the guiding field, the architecture of the deposited microcolonies is changed both under a guiding field, or upon changing the bacterial running strategy. The results for deposition rates as well as for pair correlation functions are averaged over $15$ simulation realizations. (e) The absolute value of change in orientation per unit time is calculated from an ensemble of $100$ run-and-tumble trajectories, each $60$ second long. If the run-time distribution deviates from an exponential and becomes heavy-tailed upon increasing $\alpha$, the trajectories have a lower fraction of tumble events compared to run events in a given time, and thus the absolute orientation change also reduces. This mechanism does not change the isotropy of motion in particle suspension, unlike the guiding field application where particles are directionally biased, and thus have little effect on deposition rate (b). However, the mechanism affects the pair correlation function in deposited particles (d). Error bars represent 1 standard deviation.}}
	\label{fig_fokker_planck_dNdt}
\end{figure}
The size distribution of deposited microcolonies matures relatively faster when a guiding field is applied [Fig.~\ref{fig_sizedist-combined-r00-vs-r10} (a-c) vs. Fig.~\ref{fig_sizedist-combined-r00-vs-r10} (d-f)], although, the eventual size distribution remains the same as one without the guiding field [Fig.~\ref{fig_sizedist-combined-r00-vs-r10} (f) vs. Fig.~\ref{fig_scatter} (f)]. 
{
	The deposition rate increases relatively quickly under a guiding mechanism, as depicted in Fig.~\ref{fig_fokker_planck_dNdt} (a), however, it is less sensitive to a varied running strategy (i.e., when the run-time distribution is switched from an exponential to a heavy-tailed one) [Fig.~\ref{fig_fokker_planck_dNdt} (b)].   
	The architecture of microcolonies is relatively altered both under a guiding taxis mechanism, and under a varied running strategy, suggested by the change in the pair correlation functions $g(r)$ shown in Fig.~\ref{fig_fokker_planck_dNdt} (c) and (d), for varied guiding torque and varied shape of the run-time distribution, respectively.  
	{
		The pair correlation function $g(r)$ for the non-deposited (planktonic) cell population shall look a typical $g(r)$ for (nearly) homogeneous particle suspensions, with peak near $r\geq 1$, if there is no apparent collective organization in the planktonic state. Also there is no overlap permitted so $g(r)$ should be zero up to one particle distance. However, the way $g(r)$ is computed for the deposited microcolonies is different in the present study. First a projection of the deposited cells onto the depositing plane ($z=-L_z/2$) is taken, and then the $g(r)$ for this 2D projection of cells is computed. Therefore although the particles are non-overlapping in the microcolonies due to the twitching force (Eq.~\ref{eq_twitching}), the projection can have overlaps. On the top of this, the microcolonies lie inhomogeneously on the depositing plane. Thus $g(r)$ in Fig.~\ref{fig_fokker_planck_dNdt} (c-d) exhibit fluctuations even after averaging over $15$ simulation realizations. However, the long-length scale variations are clearly visibele in Fig.~\ref{fig_fokker_planck_dNdt} (c-d), and point to alterations in the architecture of microcolonies due to guiding field [Fig.~\ref{fig_fokker_planck_dNdt} (c)] or due to change in run-time distribution [Fig.~\ref{fig_fokker_planck_dNdt} (d)].}   

	A shift in run-time distribution fundamentally changes individual particle trajectories. To understand this effect, the absolute value of change in orientation per unit time is computed in Fig.~\ref{fig_fokker_planck_dNdt} (e). The values are calculated for an ensemble of $100$ run-and-tumble trajectories, each $60$ second long. Upon increasing $\alpha$, the run-time distribution deviates from an exponential and the trajectories consist of lower fraction/ratio of tumble events to run events. Thus the average orientation change over a given time also reduces. This mechanism does not change the isotropy of motion in particle suspension, unlike the guiding field application where particles are directionally biased, and thus have little effect on deposition rate. However the change in the nature of trajectories does alter the pair correlation function Fig.~\ref{fig_fokker_planck_dNdt} (d). As the case with pair-correlation functions, the deposition rates in Fig.~\ref{fig_fokker_planck_dNdt} are also averaged over $15$ simulation realizations.}

The above insights have implications in the engineering devices, e.g. devices for magnetotaxis or galvanotaxis can be designed to precisely control or reduce the deposition of field responsive bacteria, on food surfaces for example. It should be noted that in the present study, the competition between forces that lead to the deposition of a pathogen cell onto a surface -- e.g. the electrostatic, Van der Waals, and steric interactions between the surface and the cell -- are lumped into the model for sticking probability. The benefit is the simplicity of the model and that it aims to collect the effect of numerous interactions into three floating parameters -- $P_\mathrm{c},\:P_\mathrm{s}$ and $\lambda$. The debate about competition between various aforementioned interactions between the bacterial cells and the surface is not well settled at present. In view of this, the present sticking probability model can be fitted to the experiments, and the parameters $P_\mathrm{c},\:P_\mathrm{s}$ and $\lambda$ can be deduced. Other factors influencing the initial attachment of bacterium, such as appendages on the cell's periphery, physio-chemical properties of the surface, and adhesin production~\cite{berne2018bacterial,conrad2018confined,tuson2013bacteria} are beyond the scope of this work, however, an attempt can be made to experimentally fit these effects into the aforementioned parameters in the model. 
{For instance, theoretically, the description of the cell attachment process is often attempted from the view point of different variants of the DLVO (Derjaguin-Landau-Verwey-Overbeek) theory from colloids. The point is to deduce a free energy landscape (or a potential) as the cells approach the surface, by comparing competition between electrostatic, Van der Waals, and steric forces. As we think more and more microscopically, it can be imagined that the geometric effects such as appendages on the cell’s periphery, or the roughness features on the adhering surface, might further complicate the DLVO variants. Additionally when bacterium produces adhesins for initial attachment, the situation theoretically becomes cumbersome. In such cases an empirical approach, e.g. by measuring the deposition rates and checking if the results can be reproduced using the preferential sticking probability model, can be quite useful in the absence of a theory which reliably predicts the force potentials upon cell’s approach to a surface.
}

{ 
	Possible mechanisms which can generate a guiding torque, similar to the torque modeled in the present study, also need further exploration. In addition to magnetotaxis, rheotaxis due to interstitial flows is capable of producing torque on particles, at the same time, this mechanism however can behave locally and different regions may have different directions and strengths of torque. Another relevant activity would be designing externally controllable biohybrid microswimmers with partial robotic controls. This may help generating torque on individual particles with desired direction and strength.
	Overall, present numerical and analytical results suggest that the induced asymmetrical rotational drift due to applied taxis fields, and variability in bacterial running strategies caused by a change in the shape of the run-time distributions, are important physical factors to understand the organization and early biofilm formation in collections of confined active particles.}
\vspace{-7pt}  
\section{Conclusions}
In this study, a combined numerical and analytical framework is developed to quantitatively study the aspects of early biofilm formation. One of the vital factors in the transition from planktonic to the sessile state of bacteria during early biofilm formation is the accumulation of bacteria near the surfaces. For example, self-motility with run-and-tumble dynamics in {\it E. coli} suspensions are the primary mechanisms via which the bacteria reach surfaces. It is shown in this study that a guiding taxis mechanism -- such as chemotaxis, galvanotaxis, or magnetotaxis -- can cause an asymmetric rotational drift which helps bacteria to accumulate near a surface. Conversely, by reversing the taxis field direction, the same physical mechanism can be used to drive bacterial suspensions away from the surface to reduce biomass deposition rates (e.g. using it as a food preservation technique). Exact analytical expressions -- Eq.~\ref{eq_rho_solution} to predict the density variation in accumulated cells, and Eq.~\ref{eq_soln_Ei_1} and Eq.~\ref{eq_soln_Ei} to predict the surface deposition rates -- are derived. 
{The solutions are verified with active particle simulations taking run-and-tumble statistics from multiple past experiments, including variability in cell running strategies~\cite{Note1}. 
	A change in run-time distribution from an exponential to a heavy-tailed one, as well as application of a guiding field, alters the pair correlation function in the deposited cell populations. The deposition rates are less sensitive to the change in cell running behavior but are affected significantly by a guiding field.} 
In general, the study aims to help in obtaining design parameters for external devices to alter the biomass deposition rates, and most importantly, it highlights that an induced asymmetrical rotational drift can be an important physical mechanism behind wall accumulation and organization in confined active particle suspensions.         
\vspace{-10pt}
\section*{Acknowledgements}
Present work is supported by the INSPIRE Faculty Fellowship of the Department of Science and Technology, India.

%
%
\bibliographystyle{aipauth4-2}
\bibliography{references} 

\end{document}